%% file: main.tex
\documentclass[5p,times,authoryear,twocolumn]{elsarticle}

\usepackage{booktabs}
\usepackage{mathtools}
\usepackage{setspace} 
\usepackage{etoolbox}
\usepackage{enumitem}
\usepackage{fancyvrb}
\usepackage{amssymb}    
\usepackage{subfiles}   
\usepackage{verbatim}   
\usepackage{ulem}      
\usepackage{pdfpages} 

\AtBeginEnvironment{quote}{\par\singlespacing\small}

\usepackage{xcolor}
\definecolor{ao}{rgb}{0.0, 0.5, 0.0}

\newcommand{\black}{\color{black}}

\usepackage{url}

\usepackage{breakurl}
\usepackage[unicode,breaklinks]{hyperref}
\hypersetup{
	colorlinks=true,
   	linkcolor=blue,
   	citecolor=blue,
    filecolor=blue,
   	urlcolor=blue}

\newcommand{\furl}[1]{\footnote{\url{#1}}}

\usepackage{rotating}

\usepackage{multirow}
\usepackage{tabularx}
\newcolumntype{L}[1]{>{\raggedright\arraybackslash}p{#1}}
\newcolumntype{C}[1]{>{\centering\arraybackslash}p{#1}}
\newcolumntype{R}[1]{>{\raggedleft\arraybackslash}p{#1}}

\renewcommand{\emph}[1]{\textit{#1}}

\begin{document}

\begin{frontmatter}

\title{ChatGPT, Llama, can you write my report? An experiment on assisted digital forensics reports written using (Local) Large Language Models}

\author[unil]{Gaëtan Michelet\corref{cor1}}
\address[unil]{School of Criminal Justice, University of Lausanne, 1015 Lausanne, Switzerland}
\cortext[cor1]{Corresponding author.}
\ead{gaetan.michelet@unil.ch}

\author[unil]{Frank Breitinger\corref{cor1}}
\ead{frank.breitinger@unil.ch}
 \ead[url]{https://www.FBreitinger.de}

\begin{abstract}
Generative AIs, especially Large Language Models (LLMs) such as ChatGPT or Llama, have advanced significantly, positioning them as valuable tools for digital forensics. While initial studies have explored the potential of ChatGPT in the context of investigations, the question of to what extent LLMs can assist the forensic report writing process remains unresolved. 
To answer the question, this article first examines forensic reports with the goal of generalization (e.g., finding the `average structure' of a report). We then evaluate the strengths and limitations of LLMs for generating the different parts of the forensic report using a case study. This work thus provides valuable insights into the automation of report writing, a critical facet of digital forensics investigations. We conclude that combined with thorough proofreading and corrections, LLMs may assist practitioners during the report writing process but at this point cannot replace them. 
\end{abstract}

\begin{keyword}
Digital Forensics Investigation \sep Local Large Language Models \sep ChatGPT  \sep Report Automation \sep Assisted report writing

\end{keyword}

\end{frontmatter}

\section*{Disclaimer}\label{sec:disclaimer}
It is crucial to emphasize that we do not encourage (digital forensics) practitioners to rely on LLMs to write their reports.  
Our experiments showed a non-negligible amount of hallucinations and inaccuracies which, in the real world, may lead to false allegations. 
More research is required to assess the quality, accuracy, consistency, and risks of this technology for forensic report writing.

\section{Introduction}\label{sec:introduction}

The shortage of specialized and qualified personnel combined with the increasing number of electronic devices has led to the development of a backlog in forensic labs \citep{quick2014impacts} to a point where it impacts the legal process \citep{lillis2016current}. 
To mitigate this issue, practitioners, researchers, and software vendors created tools designed to assist and automate frequent and repetitive tasks, i.e., \emph{automation}. These tools positively impact the investigation by reducing the time and effort required \citep{james2013challenges}. 
While significant progress has been made in early-phase processes such as digital evidence collection and examination (see four-phase NIST model \citep[p3-1]{kent2006guide}), assistance for reporting is still limited. This is primarily due to two reasons: 
Firstly, ``there currently is no common framework for evaluating and reporting scientific findings to mandating authorities and parties'' \citep{champod2016enfsi} and secondly, as each investigation is unique, the outcome varies significantly.

However, Large Language Models (LLMs) such as ChatGPT have gained popularity and competencies. 
Those models are designed to generate text based on a prompt submitted by the user and have proven to be efficient for different purposes, e.g., text summary, creative text generation, or text reformulation. 
While powerful, LLMs are not designed for high accuracy and suffer hallucinations, two critical problems, especially in digital forensics where a wrong assumption could lead to the sentence of an innocent person.
For instance, \citet{scanlon2023chatgpt} conducted a series of experiments ``to assess its capability across several digital forensics use cases including artefact understanding, evidence searching, code generation, anomaly detection, incident response, and education.'' 
Similarly, \citet{henseler2023chatgpt} assessed LLMs' capabilities for ``(1) writing structured queries
utilizing natural language and trace models, (2) summarizing, evaluating, and visualizing electronic communications, and (3)
analyzing search results''. 
Despite detailed experiments, the authors omit to discuss the impact LLMs may have on reporting. 
Consequently, the purpose of this case study is to explore and test the feasibility of report generation using Large Language Models (ChatGPT) and Local Large Language Models (Llama) and to answer the following research question: 

\begin{center}
    \textit{To what extent can (local) large language models assist forensic report writing?}
\end{center}

To answer this question, it is important to know about the pros and cons of large language models that have been discussed in the literature, e.g., the challenge of hallucinations. In addition, it is necessary to understand the various report sections, their content, and the sources that serve investigators as input when writing their reports. In summary, the work provides three contributions:

\begin{itemize}\setlength{\itemsep}{0pt}
    \item A thorough analysis of forensic reports to identify a consistent structure and content of these sections. 

    \item Experimental results on the quality of the generated text elements for the previously discussed sections. 

    \item An overview of the challenges faced with using Local Large Language models. 
\end{itemize}

\paragraph{Assisted vs.~automated} \cite{michelet2023automation} discussed the differences between assistance and automation and stated that ``digital forensics uses both terms (assisted and automated) as synonyms'' while other areas differentiate. For this work, we decided to use assisted as we believe that current technology can only assist practitioners but not automate the task.

\paragraph{Outline} The rest of this paper is organized as follows: First, we present the related work in the subsequent section. Sec.~\ref{sec:forensic_reporting} discusses the content and structure of forensic reports followed by a section on the \nameref{sec:generation_experiment}. In Sec.~\ref{sec:assisted_report_generation}, we summarize our findings from the conducted experiment. The three last sections include the \nameref{sec:discussion}, the \nameref{sec:limitations}, and the \nameref{sec:conclusion}.

\section{Background and related work}\label{sec:background}

This section first generally looks into advancements in automated/assisted reporting in digital forensics (Sec.~\ref{sec:automated_reporting}) followed by some general information on Natural Language Generation (Sec.~\ref{sec:nlg}).

\subsection{Automated reporting in digital forensics}\label{sec:automated_reporting}
From an academic perspective, the research outputs have been theories, concepts, and technical report generation\footnote{Note, that we make the difference between technical summaries (e.g., generated by Autopsy or Cellebrite UFED), called tool reports in this paper, and investigative reports where the latter are reports written by investigators/practitioners/examiners.}. 
\cite{van2020digital} describe the possibility of adding a reporting API in Hansken which would provide a way to generate reports, however, the project was never completed. \cite{garfinkel2012digital} suggests using the DFXML format in combination with an automated reporting tool and \cite{hayes2020adoption} mention the existence of solutions that ``analyze data, preserve it and generate a handsome report that can be used in a court of law'', but the authors do not provide an example. In their experiment, \cite{butterfield2018automated} gathered data and stored it in a database according to the ontology they developed. They explain that the stored data can be used to fill reports where the implementation is very minimal.
Some authors also present theoretical/conceptual frameworks with automated reporting capabilities: 
\cite{rughani2017artificial} describes automating the reporting process by filtering what is given by the smart analysis phase and using previously useful reports to train the smart reporting tool for each type of case. 
\cite{karie2019diverging} are exploring the use of AI to automate parts of the investigation, and in particular the possibility of assisting a user during the report generation. Their description of what needs to be included in the report feels complete due to the will to get a report that can directly be presented in court. As said, those frameworks are theories and are not implemented.

Concerning practical work, \cite{jain2014comparative} claim a tool with an automated generation of reports module that takes place after the investigation module. However, the authors do not provide the tool, a sample report, or a detailed description.
\cite{farrell2009framework} presents his vision for an automated report generation framework using PyFlag. While he describes it as a report, to us this is a summary of findings that can be used by an investigator for further consideration. For instance, summaries (separated by user account) include information such as the most visited websites, a list of images (including the size, the name, and the link to the image), the number of files for each file type, etc.

The private sector also addressed the topic and included reporting on their products. Autopsy\furl{https://www.autopsy.com/} or Cellebrite UFED Physical Analyser\furl{https://cellebrite.com/en/physical-analyzer/} provide a feature to generate a ``report''. Those reports are of a technical nature and provide a summary/overview of elements found on the analyzed device. The results are presented based on the analysis and originate from the files of interest based on the FileSystem tree structure. This automatically generated ``report'' could fit into the documentation log of the investigator, but cannot directly be used ``as is'' in a forensic report. 
Recently, some of the industry corporations like SecurCube\furl{https://securcube.net/forensics-report/} or MSAB\furl{https://www.forensicfocus.com/reviews/xamn-report-builder-from-msab/} started to propose modules that could create different parts of forensic reports based on templates defined by the user or the tool (more often by the tool), along with some user-inputted data. The generated passages are a good start but still fairly straightforward and products of commercial software.

\subsection{Natural Language Generation}\label{sec:nlg}
The automated generation of text has been explored in the area of Natural Language Generation (NLG). This sub-domain of Natural Language Processing targets the generation of text. The most common model for NLG implementation requires the accomplishment of six tasks \citep{reiter_dale_1997}. For each one of them, different methods and techniques can be used, for example template-based, rule-based, or learning-based methods \citep{gatt2018survey}. Each of these techniques does not provide the same amount of control over the generation of the text, in particular stochastic approaches such as the learning-based methods. Using the previously mentioned model, different authors explored the application of NLG to reports and automated their generation in several areas: health care \citep{cawsey1997natural}, mammography \citep{hoogi2020natural}, weather forecasts \citep{yao1998system}, air quality \citep{wanner2007measurements}, etc. An overview of the report generation topic was provided by \cite{wanner2010report}.

\section{Content and structure of forensic reports}\label{sec:forensic_reporting}

Answering the research question (Sec.~\ref{sec:introduction}) requires an extended knowledge of forensic reports wherefore this section raises a sub-research question:

\begin{center}
\textit{Which sections or elements in a forensic report are suitable for creation through the use of (local) large language models?}
\end{center}

To respond to this question, we perform a comprehensive analysis of forensic reports. That is, we provide a detailed exploration of the report structure, textual elements, data prerequisites, and an assessment of their LLM-potential, i.e., which parts may be assisted by LLMs. 
Here, the term (textual) \emph{element} is used as a generic term for content found in sections and paragraphs, e.g., text, tables, images, listings, etc. Alternative terms could be report component or part. 
Specifically, we performed three key steps:
\begin{description}\setlength{\itemsep}{0pt}
    \item[Defining a consistent report structure:] The initial step involves defining a standardized structure for forensic reports. A uniform structure ensures that generated content is coherent and aligns with the expectations of report recipients.

    \item[Comprehending section contents:] This looks at the purpose and content of each report section to understand the specific objectives and information they need to convey.
    
    \item[Identifying information sources:] The process of generating content using LLMs necessitates a clear identification of where the input data (source) comes from including its format.
    
\end{description}

\subsection{Types of forensic reports}
According to \cite{horsman2021different}, there are three types of forensic reports, namely technical reports, investigative reports, and evaluative reports. 
The technical report represents a grouping of facts describing the analyzed data, providing information such as the quantity, location, or type of data. 
The second type of report contributes to another layer of information by providing potential explanations for the findings. 
Lastly, evaluative reports present a probabilistic approach that tries to establish the strength of the evidence by evaluating the likelihood of their finding under a set of competing hypotheses (here two competing explanations provided in the investigative report for example).
In addition, there are tool reports, i.e., reports that are generated by forensic software suites such as Autopsy\furl{https://sleuthkit.org/autopsy/docs/user-docs/4.20.0/reporting\_page.html} or Cellebrite UFED\furl{https://cellebrite.com/en/the-complete-guide-to-generating-reports-in-physical-analyzer-a-step-by-step/}. The findings are represented using different types of data structures such as listings or tables (one usually finds less free text).

\emph{Remark:} For the remainder of the paper, we will use the terms \emph{tool report} and \emph{forensic report}. The latter is the document written by an expert and submitted to the court and does not align with \cite{horsman2021different}'s terminology.

\subsection{Input data source}\label{sec:input_data_source}

In addition to the \emph{tool report} mentioned in the previous section, an examiner relies on three additional sources to write the forensic report:

\begin{description}
    \item[Mandate:] The mandate is written by the prosecutor and includes details about the case such as the context that led to the investigation, the pieces sent for analysis, and investigative questions.

    \item[Lab log:]  The lab log is the examiner's logbook and includes details about the steps taken, observations, techniques, etc. It can be long/comprehensive and often written in natural language (bullet points).
    
    \item[Knowledge and experience:] Of course, while writing the report, the examiner utilizes knowledge and experience from the case itself as well as the past to draw conclusions. It is impossible to access this information if it is not written in the lab log. 
\end{description}

\subsection{Methodology}

First was the identification of a standardized layout of reports, i.e., common sections found in reports.

\begin{enumerate}\setlength{\itemsep}{0pt}
    \item Assessment of guidelines: we conducted a review of some existing guidelines and teaching materials to identify commonalities in report structuring, e.g., by \cite{Pollitt2014forensic,garrie2014digital,casey2011digital} as well as considered the structure that we apply within our institution. 
\end{enumerate}

For each identified section, we were then interested in the purposes, content, internal structure, input data source, and LLM-potential which was researched as follows: 
\begin{enumerate}\setlength{\itemsep}{0pt}
    \setcounter{enumi}{1}
    \item Empirical analysis of sections: The authors reviewed approximately one hundred student-generated reports\footnote{The reports were written as homework assignments for two courses. Course one: undergraduate students write their first and second digital forensics reports with a focus on USB-stick acquisition and computer analysis. Course two: graduate students write two reports on file-recovery and mobile analysis.} created during coursework to learn about the \textbf{purpose and content} of these sections. In addition, we did a structural analysis, i.e., we identified the \textbf{structure and elements} that make up the section.

    \item Input source identification: This phase focused on identifying the specific data requirements needed to produce each report element, i.e., which \textbf{(input) data source} may be used to write this section.
    
    \item LLM-potential rating: Lastly, based on a synthesis of the findings from the previous steps, we provide an initial assessment of the chances that this section (or parts thereof) can be generated by an LLM which is based on the structure variability and data availability: a good data availability and low variability in the section results in a high \textbf{LLM-potential} and vice-versa. 
\end{enumerate}

\subsection{Results and discussion of forensic reporting}\label{sec:forensic_report}
While we acknowledge that every investigation is different and thus every report is unique, there are commonalities among different reports in terms of structure and content. This is an essential finding as otherwise, automation would not be possible at all.

As the outcome of \emph{the assessment of guidelines} (step 1), we identified six core sections (exact wording for the section names varied): 
(1) Introduction,
(2) Items received,
(3) Methodology,
(4) Results,
(5) Discussion, and
(6) Conclusion. 
Given these sections, a summary of our findings (steps 2 to 4) is provided in Table~\ref{tab:summary_reports}. The details are highlighted in the upcoming subsections organized by the characteristics. 
Note that the ``Items received'' section was sometimes included as a subsection of the ``Introduction'' or ``Results'' section. Given that it was a complete and separate section on the majority of the reports, we decided to consider it as a whole section in this study.

\begin{table*}[th]
{\centering
\resizebox{\textwidth}{!}{%
\begin{tabular}{c|c|c|c|c|}
\cline{2-5}
                                              & \textbf{Purpose and Content}                                                                                                                                                                                                                                                                                        & \textbf{Structure and Elements}                                                                                                                                                                                                     & \textbf{Input data source}                                                                                                                                           & \textbf{LLM-potential} \\ \hline
\multicolumn{1}{|c|}{\textbf{Introduction}}   & \begin{tabular}[c]{@{}c@{}}Provides a summary of the mandate and the investigation context, \\ includes crime description, suspect(s), investigator(s), \\ transmitted items, and the prosecutor questions\end{tabular}                                                           & \begin{tabular}[c]{@{}c@{}}Text that usually follows the mandate structure \end{tabular} & Mandate and Lab Log                                                                                                                                                  & High                   \\ \hline
\multicolumn{1}{|c|}{\textbf{Received Items}} & \begin{tabular}[c]{@{}c@{}}Description of seized items including 
characteristics: \\ size, hash (if forensic image), or physical condition (if device) \end{tabular} & \begin{tabular}[c]{@{}c@{}}Combination of texts, tables,\\ and lists describing each item\end{tabular}                                                                                                                              & \begin{tabular}[c]{@{}c@{}} Mandate, Lab Log and Tool report (note that\\ this last source only presents partial data) \end{tabular}                                                                                                                                                   & High                   \\ \hline
\multicolumn{1}{|c|}{\textbf{Methodology}}    & \begin{tabular}[c]{@{}c@{}}Details analysis procedure, i.e., steps followed (incl.\\justification) and tools used (incl.~versions) \\ \end{tabular}                                                                                                                                                      & \begin{tabular}[c]{@{}c@{}}Text or list of taken steps/applied tools \\ in chronological order\end{tabular}                                            & \begin{tabular}[c]{@{}c@{}}Literature, and Lab Log\end{tabular} & High                   \\ \hline
\multicolumn{1}{|c|}{\textbf{Results}}        & \begin{tabular}[c]{@{}c@{}}Provides an overview of the results,  e.g., a list \\ of every artifact of interest identified, and their characteristics\end{tabular}                                                                                                                                   & \begin{tabular}[c]{@{}c@{}}Combination of texts, tables,\\ and lists with varying structure \end{tabular}                                                              & Lab Log and Tool report                                                                                                                                             & Medium*
\\ \hline
\multicolumn{1}{|c|}{\textbf{Discussion}}     & \begin{tabular}[c]{@{}c@{}}Discusses the meaning of the results in the context of the \\ investigation, and the limits of the undertaken analysis$^{\dagger}$\end{tabular}                                     & \begin{tabular}[c]{@{}c@{}}Text with varying structures for each mandate's \\ question (the evaluation usually comes at the end)\end{tabular}                                                                         & \begin{tabular}[c]{@{}c@{}}Investigator knowledge, experience, and opinion\\ (parts of the data may be in the lab log)\end{tabular}                                 & Low                    \\ \hline
\multicolumn{1}{|c|}{\textbf{Conclusion}}     & Summarizes the important elements of each prior section                                                                                                                                                                                                                                                             & \begin{tabular}[c]{@{}c@{}}Text following the overall structure of the report \end{tabular}                    & Prior elements                                                                                                                                                       & Medium-Low             \\ \hline
\end{tabular}%
}}
{\footnotesize * Low for the whole section but high for various components within the section.\\ 
$^{\dagger}$ An evaluation of the results under the light of each hypothesis is also present if the report is evaluative.\\
}
 
\caption{Summary of findings for the forensic report analysis.} \label{tab:summary_reports}

\end{table*}

\subsubsection{Introduction section}

\begin{description}\setlength{\itemsep}{0pt}
    \item[Purpose and content:] The purpose of the Introduction is to present the context of the investigation, such as the crime investigated, the transmitted items, the persons or entities that are involved (including the mandated investigator), and the investigative questions asked by the prosecutor. 
    
    \item[Structure and elements:] This section usually follows the mandate and most frequently starts with a description of the potentially committed crime and the name of the suspects, followed by a summary of the seized items to be analyzed, the name of the investigator mandated by the prosecutor, and the investigative questions to answer. It is often in paragraph format.
    
    \item[Input data source:] The content for this section originates primarily from the mandate (of course it should also be found in the lab log). This information may be complemented with additional data, e.g., personal details of the investigator.
    
    \item[LLM-potential:] As this part is foremost a summary and reformulation of the received mandate, this section is considered a good candidate for generation.
\end{description}

\subsubsection{Items Received section}

\begin{description}\setlength{\itemsep}{0pt}
    \item[Purpose and content:] This section provides information about the items that were sent to the investigators for analysis (compared to the Introduction, it provides more details) where an item may be a physical device or a forensic image. The section highlights aspects such as the size of the storage, the hash value, or the procedure used to acquire the data. For physical devices, a description of the physical state is added, e.g., potential damages.
    Photos are usually not placed here but can be found in an appendix.

    \item[Structure and elements:] The information is frequently presented in a table, a list, or in short paragraphs describing each device/image. 
    
    \item[Input data source:] Content for this section originates from the mandate and the investigator's lab log which should include descriptions of the devices. Note, that tool reports normally only contain information related to device images.
    
    \item[LLM-potential:] Given the availability of the information and the structure of the text/table, this section has a high LLM-potential.
\end{description}

\subsubsection{Methodology section}

\begin{description}\setlength{\itemsep}{0pt}
    \item[Purpose and content:] This section includes a comprehensive breakdown, rationale, and justification for each step adopted during the investigation, along with a summary of the tool(s) used to facilitate each of these steps.
    
    \item[Structure and elements:] The organization of the content is a straightforward and chronological description of each step throughout the investigation. The utilized tools are frequently displayed in a table or footnotes.
    
    \item[Input data source:] Various methodologies (models, frameworks) have been discussed in the literature and are well-known by practitioners (e.g., discussed during training). LLMs should be aware of these models during their training. 
    The exact steps (details) should be captured in the lab log including the list of tools, their versions, and purpose.

    \item[LLM-potential:] Based on these aspects, this section is considered to be a valid candidate for LLM.
\end{description}

\subsubsection{Results section}
\begin{description}\setlength{\itemsep}{0pt}
    \item[Purpose and content:] All relevant results of the analysis carried out are presented. The various artifacts that can help answer the mandate's questions are also detailed. 
    
    \item[Structure and elements:] The structure varied significantly between sample reports confirming that a logical organization is complex. The artifacts and evidence were sometimes presented by type, sometimes chronologically, and sometimes based on the question they helped to answer. Moreover, the section frequently utilizes various elements such as text, tables, lists, or diagrams. 
    
    \item[Input data source:] All the necessary context is available both in the tool report and the lab log. The examiner's experience may impact this section.
    
    \item[LLM-potential:] Given the complexity, generating this whole section at once is considered to be impossible (at the time of writing). However, the generation of subsections/elements for various artifacts is deemed possible, e.g., a summary of an artifact.
    
\end{description}

\subsubsection{Discussion section}
\begin{description}\setlength{\itemsep}{0pt}
    \item[Purpose and content:] In this section, the limits of the analysis are shown and the results are put into context. It is also the place where the Bayesian evaluation of the results given each hypothesis takes place if the mandate is asking for such an assessment\footnote{In this case, the section is often named Evaluation instead of Discussion.}. It allows the reader to better comprehend the significance of the findings and to be aware of possible errors caused by the instruments, the methods of analysis, or the interpretation of the results.
    
    \item[Structure and elements:] The structure of this section varies considerably among the reports. However, a commonality is that each mandate question is discussed where, for each question, the limits/considerations are mentioned as well as the interpretation of the evidence. The evaluation is usually the last element to be presented.
    
    \item[Input data source:] The content comes from the experience, knowledge, and opinion of the mandated investigator/expert. Much of it will be in the lab log, but also the head of the examiner. In contrast, it cannot be found in the tool reports. 
    
    \item[LLM-potential:] The limited availability of the required data and the complex organization of the text make this element of the report difficult to automate. 
\end{description}

\subsubsection{Conclusion section}

\begin{description}\setlength{\itemsep}{0pt}
    \item[Purpose and content:] The final section reiterates the important elements of the investigation and rounds the report.
    
    \item[Structure and elements:] The structure frequently follows the overall structure of the report, starting with elements of the Introduction, then the presentation of the Received Items, the Methodology, the important Results, and eventually the core discussion points (the received items and methodology are sometimes also presented just before the results). 
    
    \item[Input data source:] The input for this section is the report itself with a focus on the Introduction, the Results, and the Discussion.
    
    \item[LLM-potential:] This section relies on experience but also an understanding of the complete report which can be rather long. One may attempt to copy and paste the full report and demand a conclusion but the LLM may not accept such large inputs. However, the LLM may help with parts of it.

\end{description}

\subsection{Summary}
This section aimed to identify the different sections, content, and elements of reports that, based on their content and structure, are good candidates for being drafted by LLMs. Out of the six identified sections, the potential was considered high for the Introduction, Items received, and Methodology as well as a medium potential for the `Results'.
In comparison, the complexity of the Discussion and the Conclusion is estimated to be higher. 
For the remainder of this article, we decided to not consider the latter two further and focus exclusively on the first four.

\section{Experimental setup for assisted report generation}\label{sec:generation_experiment}
To test the feasibility of assisted report generation, we decided to conduct an experiment where we will work on a fictive case, have a mandate, take notes in the form of a lab log, and then eventually have to deliver a forensic report. This section summarizes the setup before discussing the results in Sec.~\ref{sec:assisted_report_generation}.

\subsection{Large Language Models}
The experiment uses two models: Llama-2 which runs locally and ChatGPT-3.5 which runs online.

\paragraph{Llama-2} Llama is a popular large language model developed by META and is free to use for research purposes. 

The model is available in several versions with varying characteristics. For this study, we considered the number of parameters and the quantization method which both directly influence the quality of the generated text. 
Precisely, the more parameters the model has and the more precise the quantization, the better the quality of the output text should be. 
The price for this quality improvement is the computing power and/or time required to create the text. A balance must therefore be struck between trying to optimize the quality of the text while ensuring that the text is produced within a reasonable time frame. We chose a quantization method (q4\_1) and then tested three versions of Llama-2 with a varying number of parameters (7B, 13B, and 70B) on our workstation\footnote{Everything was set up on an average desktop: a Windows 10 64-bit operating system, an AMD Ryzen 5 3600X 6-Core (3.80\,GHz) Processor, and 16\,GB of DDR4 memory.}.
The text quality of the 7B model was poor and the required time for the 70B model was too long.
Consequently, the Llama-2-13B \black original version (not fine-tuned) with a q4\_1 quantization (downloaded on the HuggingFace page of ``the bloke''\furl{https://huggingface.co/TheBloke}) was chosen for this experiment. 

To interact with the model, Koboldcpp\furl{https://github.com/LostRuins/koboldcpp} was used which is an easy-to-use application for models in the GGML and GGUF format\footnote{GGML and GGUF are file formats that are used to store models for inferences, especially in connection with language models such as GPT (Generative Pre-trained Transformer).}. With Koboldcpp, the model can be accessed through a web GUI (similar to ChatGPT) on port 5001 or through an API allowing to automate the sending of requests and the reception of the answers. 

\paragraph{ChatGPT-3.5} As our system lacked the computing power to run large models, we anticipated a limited quality of Llama. Consequently, we decided to also evaluate GPT-3.5 via the Open\-AI ChatGPT interface. Nevertheless, we consider the local model of utmost importance as this better reflects the real world: Investigators likely lack the necessary resources to deploy sophisticated language models like GPT-3.5 in their local environment. Furthermore, utilizing online models is not feasible due to the confidential nature of the data.

\subsection{Fictive case description}\label{sec:case-presentation}
The fictional case, created for a cell phone analysis course evaluation, revolves around incidents that occurred in 2019. It involves two suspects, Mr.~Pressive and Mr.~Sforza, who are believed to be associated with the theft of the American Declaration of Independence. Two devices (evidence) were collected: A Samsung Galaxy S6 Edge and an iPhone 6.

The investigator was tasked with answering the following questions based on the collected evidence (a copy of the mandate is given in \ref{app:mandate}):
\begin{itemize}\setlength{\itemsep}{0pt}
    \item Where was Mr.~Sforza in January and February 2019?
    \item Where was Mr.~Pressive in January and February 2019?
    \item Did Mr.~Sforza and Mr.~Pressive meet during this period?
    \item Did Mr.~Sforza and/or Mr.~Pressive take part in any illegal activities during this period?
    \item In particular, are Mr.~Sforza and/or Mr.~Pressive implicated in the theft of the US Declaration of Independence?
\end{itemize}

To answer these questions, location-related data, and communication records are crucial. Key artifacts for analysis include accounts, communications logs, location data, and pictures.

\subsection{Working on the case}
While we are familiar with this (low complexity) case, the experiment aimed to be as realistic as possible. It started with us receiving the forensic images along the mandate. Then, we utilized a forensic workstation to conduct the analysis. 

\subsubsection{Extraction and processing}
After receiving the devices (evidence), we performed the following extractions:
\begin{itemize}\setlength{\itemsep}{0pt}
    \item On the Samsung Galaxy S6 Edge: logical and physical extractions were performed.
    \item On the iPhone 6: logical and file system acquisitions were performed, and Telegram messages were manually captured. \black

\end{itemize}

Next, we processed the images using Cellebrite UFED Physical Analyzer (version 7.3.0.75) which resulted in two tool reports (PDF and UFDR\footnote{UFDR stands for Universal Forensic Extraction Device Report and is a format proposed by Cellebrite. It is an interactive format that allows browsing (filtering, searching, etc).}). 
For simplicity, we only utilized the PDF report which we accessed using the Acrobat Reader.  
The report is about 3600 pages long and contains a considerable amount of case-irrelevant details. Thus, case-relevant parts were copied into the lab log with minor modifications. As an example, general device characteristics of the smartphones were reduced as shown by Table~\ref{tab:samsung_lab_log} (lab log) and Table~\ref{tab:samsung_tool_report} (tool report).

\begin{table}[th]
\centering \footnotesize

\begin{tabular}{ll}
\toprule
\multicolumn{2}{c}{\textbf{Samsung Galaxy S6 Edge} } \\
\midrule
Detected Phone Vendor   & Samsung \\ 
Detected Phone Model    & SM-G925F \\ 
OS Version              & 6.0.1 \\ 
MAC address             & AC:5F:3E:73:E3:78 \\ 
Time Zone               & America/New\_York \\ 
\bottomrule
\end{tabular}
\caption{Table describing characteristics for the Samsung Galaxy S6 Edge in the lab log.} \label{tab:samsung_lab_log}
\end{table}

\begin{table}[th]
\centering \footnotesize

\begin{tabular}{lll}
\toprule
\textbf{Name}           & \textbf{Value}        & \textbf{Source} \\ 
\midrule
Detected Phone Vendor   & Samsung               & build.prop: 0x2AC \\ 
Detected Phone Model    & SM-G925F              & build.prop: 0x292 \\ 
OS Version              & 6.0.1                 & build.prop: 0x132 \\ 
MAC address             & AC:5F:3E:73:E3:78     & .mac.info: 0x0 \\
Time Zone               & America/New\_York     & persist.sys.timezone: 0x0 \\ 
\bottomrule
\end{tabular}%

\caption{Table describing characteristics for the Samsung in the tool report} \label{tab:samsung_tool_report}
\end{table}

\subsubsection{Artifacts of interest}
Answering the first two questions requires location-related artifacts. Our starting points were the WiFis\footnote{Note that the Samsung device did not contain any relevant WiFi location data.} that the device connected to (in particular the SSID and BSSID), as well as the GPS EXIF metadata of taken pictures. 
The visual content of the pictures and text messages (communication exchanges) are used as supportive arguments. 
    
The third question can be addressed by correlating the extracted locations and timestamps of the two devices. Meeting points found in the communications and pictures of common landmarks (e.g., monuments or popular places) are also analyzed and findings are written to the lab log. 
    
Concerning the last two questions, we concentrated on the messages exchanged between the suspects. To draw conclusions about the activities of Mr.~Sforza and Mr.~Pressive, we examine the messages that were sent between the two phones via WhatsApp, Telegram, and Email. Findings are then related to the previously identified locations. Lastly, we searched the browser history (Chrome, Safari) for additional evidence.

\subsection{(Input) Data formats and transformation}
The lab log can be seen as a condensed version of the tool report that focuses on relevant details and improved readability. To accomplish that, some details were omitted, or replaced (extended) with other information. Specifically, the following things were done: 
\begin{itemize}
    \item EXIF information of pictures includes GPS coordinates. Converting these coordinates into addresses or locations using Google Maps makes it easier to grasp them. Consequently, one may replace the Latitude and Longitude with a related location and format it; compare Fig.~\ref{fig:ufed} and Fig.~\ref{fig:log}.
    \item Cellebrite is currently not generating descriptions of the pictures stored on the device. Thus, we manually added brief descriptions of images to the lab report.
    \item Lastly, the Linking between traces was added manually, e.g., between locations (GPS), pictures, and conversations. 
\end{itemize}

\begin{figure}[ht]
\scriptsize
\begin{verbatim}
    # Source Position Info Confidence Category Deleted
    1 (38.907500, -77.072778) Name: 20190214_143344.jpg
    Hour: 02/14/2019 14:33:44
    Source File: USERDATA
    (ExtX) /Root/media/0/DCIM/Camera/20190214_143344.jpg : 
    0x364 (Size : 4894559 octets)
    Media Locations
    2 (38,907500, -77,072778) Name: 20190214_143342.jpg
    Hour: 02/14/2019 14:33:42
    Source File: USERDATA
    (ExtX) /Root/media/0/DCIM/Camera/20190214_143342.jpg :
    0x364 (Size : 4239509 octets)
    Media Locations
    3 (38.906944, -77.072500) Name: 20190214_142901.jpg
    Hour: 02/14/2019 14:29:01
    Source File: USERDATA
    (ExtX) /Root/0/DCIM/Camera/20190214_142901.jpg : 
    0x364 (Size : 3756708 octets)
    Media Locations
\end{verbatim}
    \caption{Several locations found in pictures from the Cellebrite tool report (copy of tool report).}
    \label{fig:ufed}
\end{figure}

\begin{figure*}[ht]
\scriptsize 
\begin{verbatim}
Name	               Time	               Category	       Latitude	        Longitude	         Related Location
20190214_143344.jpg	14.02.2019 14:33:44	Media Locations	38,9075	         -77,0727777777778	 Healy Hall, 37th St NW, Washington, District of Columbia 
20190214_143342.jpg	14.02.2019 14:33:42	Media Locations	38,9075	         -77,0727777777778	 Healy Hall, 37th St NW, Washington, District of Columbia
20190214_142901.jpg	14.02.2019 14:29:01	Media Locations	38,9069444444444 -77,0725	          Northwest Washington, Washington, District of Columbia
\end{verbatim}
    \caption{Same locations as in Fig.~\ref{fig:ufed} but formatted for better readability and with an additional field for human-readable location (copy of lab log).}
    \label{fig:log}
\end{figure*}

The lab log primarily contains tables summarizing relevant information. This is different from the tool report which also uses bullet points.

\section{Assisted report generation using LLMs}\label{sec:assisted_report_generation}
This section assesses how, based on the various input data (extracted from different data sources, Sec.~\ref{sec:input_data_source}), the sections of the forensic report can be filled. 
This will confirm or disprove our \emph{LLM-potential} (discussed in Sec.~\ref{sec:forensic_reporting}) and thus give a first answer to the research question raised in the introduction. 

Note that we limited the experiments to a subset of the report, i.e., we are not generating the complete final report but do sampling within each section.

\subsection{Input text for the LLMs}
For our experiment, we tested various input text formats: the mandate consists of paragraphs, the lab log (MS Word) consists primarily of tables and the tool report (PDF) is a mixture of bullet points and tables as well as lots of irrelevant details. All serve as input text for the LLMs. Note, as we use copy+paste formatting will be lost when copying it into the LLM web interface. 

Although we do not expect that the lost formatting will have a significant impact on the output quality, we decided to create another input text format with the LLM prompting in mind. 
This option is a copy of the lab log word table that was filtered to only keep information relevant for the report generation, and formatted in the CSV format.

\subsection{Procedure for assisted reporting}
The report sample text will be generated with both Llama-2 and GPT-3.5, using 36 prompts (per LLM) using the input data outlined in Table~\ref{tab:summary_reports}.

\paragraph{Inputs} For the first two sections (introduction, received items), the mandate served as an input. For the methodology, the methodology section of the lab log was used. For the two subsections of the results (communications, locations), we utilized the three different input formats, i.e., the tool report, the lab log, and the CSV-formatted data.

\paragraph{Prompts} Each LLM was queried four times for each input-section combination, i.e., 4 for the introduction, 4 for the received items, 4 for the methodology, 12 for the communication summary, and 12 for the location summary. The requests were kept simple, e.g., `I need your help to write the section $<$SECTIONNAME$>$ of a forensic report' combined with the input data. For the results-subsections, requests made to the LLM asked for a ``summary'', and for a ``day-by-day summary'' (e.g., of the conversations). 
The position and formulation of the request changed slightly with the request sometimes being located at the beginning of the prompt, and sometimes at the end. 
Note, that it was decided to reset the session after each prompt to prevent the influence of previous interaction. We believe this reflects the real world as investigators likely will not submit multiple queries for the same report element in a row.

\paragraph{Quality} The quality assessment of the generated text was done manually where emphasis was given to the accuracy of the content and its structure. \black

\subsection{Detailed result for one of the queries/prompts}
Here is a more detailed description of one of the created prompts, designed for the generation of the Introduction section, and the text returned by Llama. Note that as previously stated, the content of the four queries for a specific Input data source is the same, but the form of the input data, or the formulation of the request can change.

The first part of the prompt provides a copy of the mandate in a structured way, including the date of reception, the description, and the mandate (some parts of the text were removed to improve the readability and reduce the size of the example):
\begin{small}
\begin{verbatim}
    "Date: 01.10.2023
    Description: Following an international arrest 
    warrant, two individuals, Mr Sforza and Mr 
    Pressive, were arrested in Lausanne on (...)
    Mandate: Using the information provided, we ask
    you to answer the following questions: i. Where
    was Mr Sforza in January and February 2019? (...)
    We request that you answer these questions
    before the 12th of October 2023 
    Note: The person mandated is the investigator X"
\end{verbatim}
\end{small}

The second part is the request, specifying the report part needed, and the elements that should be present in the section.

\begin{small}
\begin{verbatim}
    Can you summarize the previous text and write the 
    intro of a forensic report for me? I need important
    elements of the description, the mandate, the 
    questions asked (all of them), and the investigator
    of the case!
\end{verbatim}
\end{small}

This prompt was submitted to the two models, and here is the result obtained through the Llama Model:

\begin{small}
\begin{verbatim}
    The two suspects were both arrested in Lausanne 
    on suspicion of stealing the Declaration of 
    Independence. The authorities do not have photos
    of Mr.Sforza or Mr. Pressive. The phone of Mr.
    Sforza was stolen by Italian police. The phones
    of both men will be analysed. Mr. Sforza was in
    Iraq until March 2020 and Mr. Pressive was in
    Switzerland until early Februry 2019. I can't
    find evidence of any illegal activity by either.
\end{verbatim}
\end{small}

The result obtained for each prompt was saved and later analyzed.

\subsection{Results Summary}

This section includes a summary of the results, highlighting the discovered benefits of LLM-assisted report writing, as well as the associated challenges and issues. A complete description of the results is provided along with the result's data, including the prompts created and the report elements generated which can be accessed here: \url{https://github.com/Michelet-Gaetan/ChatGPT_Llama_can_you_write_my_report}. 

As outlined earlier, each section comes with its challenges and requires different inputs. We therefore separated the following sections by the names of the generated elements.

\subsubsection{Positive Outcomes}
\paragraph{Introduction} The task of the LLM here is to extract relevant details from the mandate (in our case, additional elements where required, for example, the name of the mandated investigator) and present them in a logical order. This was consistently achieved by ChatGPT, which produced various accurate and complete introductions that could fit into a report with minor adjustments. Llama's performance, on the other hand, was worse but it still returned text that could be integrated into a forensic report after proofreading and corrections. These results are encouraging, and even if some of the report introductions created by Llama could not be used, both models generated at least one element of sufficient quality. Consequently, our assumption of the LLM-potential of the introduction is true.

\paragraph{Items received} ChatGPT texts were accurate and complete, making them suitable for report inclusion after minor adjustments. Llama's outputs varied in quality, and while they showed potential, they often lacked accuracy and correctness, necessitating extensive proofreading and corrections. Note, that sometimes this may not be necessary as it may be directly copied from the mandate or log. Overall, we are convinced of the expected automation capability.

\paragraph{Methodology} This section is an overview of the various steps taken to conduct the analysis, along with their justification and the tools used to accomplish them. The model must therefore parse the input data, here a list of achieved steps along with their purpose and a list of the used tools, and present them in a logical order (most of the time chronological). One of the challenges is to correctly explain the goal of each step, and mention the tool used to achieve it, something ChatGPT was able to do. In general, the outputs generated by the online model were accurate and complete. 
An interesting behavior of the model was that a significant amount of (correct) auxiliary details were added when generating this report section with the GPT model. This may be explained by the fact that ChatGPT was trained on a dataset containing several methodologies of mobile forensic analysis. 
Llama also created some elements with acceptable completeness, but none of the text was sufficiently accurate (this will be discussed in the next section). 
In summary, ChatGPT was capable of generating first drafts that, once proofread and corrected, could be introduced in a forensic report. This matches our expectations regarding this section's automatability.

\paragraph{Results} We limited the text generation to two pieces of evidence namely the conversations and the GPS locations: 

For \emph{conversations}, it is essential to identify/extract important messages from the input and summarize them. 
We examined the behavior of three input formats: a copy from the tool report, a copy from the lab log, and a filtered/reformatted version of the lab log (see Fig. \ref{fig:ufed_chat_messages}, \ref{fig:log_chat_messages}, and \ref{fig:csv_chat_messages}, respectively). 
Text generated by ChatGPT had a wide range of quality with varying levels of accuracy and completeness. Several of them are considered of sufficient quality to serve as a first draft. An interesting observation was that the quality of the outcome was not impacted by the input format (two of them contained large amounts of irrelevant data).
On the other hand, Llama's performance was insufficient concerning accuracy and completeness. 

For the \emph{locations}, the goal is to extract and group common locations and provide a summary of the regions in which the device was located during the period of interest. To do this, the model must parse the locations (GPS coordinates or locations/addresses) from the input data, group them spatially or chronologically, and present these groups. The tested sources of data are similar to the one used in the previous paragraph. Note that here, the tool report only provides GPS coordinates, while the table extracted from the lab log contains the addresses. 
ChatGPT generated mostly accurate summaries, but sometimes incomplete. Once again, Llama did not produce any text of sufficient quality. With adjustments and proofreading, the texts generated by ChatGPT could be integrated into a forensic report.

The different queries did not have a significant impact on the generated text quality. We only observed minor changes when asking for ``summary'' and ``day-by-day summary'' for the messages and locations. 
Most of the time, only structural differences were observed (may be used to optimize for personal or institutional output preferences). For the conversation summaries, the day-by-day version was overall slightly more accurate and easier to understand, but less complete. For the location summary, no difference other than structural was observed.

Lastly, some of the prompts included French words due to the default language setting when generating the report (which were removed from the examples listed in this article). This could have impacted the quality of the generated texts, but no French word or quality diminishing was observed with the tool report input data.

\subsubsection{Challenges and Issues}

\paragraph{Introduction \& Items Received} Llama generated texts with inaccuracies and hallucinations. While some showed potential, they often lacked accuracy and correctness, necessitating extensive proofreading and corrections. ChatGPT's generated texts, while occasionally incomplete or inaccurate, included additional information not found in the input data. This required restructuring to ensure the suitability of the report.

\paragraph{Methodology} Outputs had similar problems. The elements created by Llama had a high amount of inaccuracies (in particular with the tools version) and hallucinations (hallucinations were plausible for a non-digital forensics specialist, which is more problematic than before). It is possible to integrate these texts into the report and proofread/correct it. However, it is probably more efficient to write the methodology oneself.

\paragraph{Results (i.e., conversations and locations)} Llama's limitations became obvious when trying to generate texts based on the tool report and lab log data. The time taken to create the text was approximately 30 minutes, and we considered it infeasible in practice. On the other hand, when texts were generated from the modified input data, they were of poor quality, often incomplete, and presenting a significant amount of inaccuracies. ChatGPT's texts were often incomplete, but most of the time accurate. The limits of the GPT model were reached with the location summaries generated based on the tool report. The data was only constituted of GPS coordinates, and the model was not able to transform them into human-readable locations, before grouping them. Moreover, signs of hallucination were present, with for example the presence of GPS coordinates that were not part of the input data. Tests were run to see if GPT-3.5 could accurately provide an address for a given Location given its GPS coordinates, and that was the case.

\section{Discussion}\label{sec:discussion}
In summary, ChatGPT generally demonstrated better performance overall, but thorough proofreading and adjustments were consistently needed for both LLMs to ensure the suitability of their outputs for report integration. Llama faced challenges with time constraints and data complexity, while ChatGPT occasionally introduced inaccuracies and additional content that required careful review.

\subsection{Local LLM vs.~LLM}
The quality of the texts generated by the two models was significantly different. The elements created by ChatGPT were systematically more accurate, clear, and complete than the ones generated by Llama. This can be explained by the difference in the model's size and capabilities: due to the computational limits of our workstation, the 13B version of the Llama model was the biggest one that could be run. Tests were undertaken with Llama70B, but this version took an unreasonable amount of time to generate a piece of text (over three hours for 500 tokens). The GPT-3.5 model used by ChatGPT is significantly larger and is therefore of better quality. Another important difference that was yet not mentioned is the amount of time taken to generate the requested elements, which is once again related to the computational capabilities and the size of the models. The average duration was 10 seconds for ChatGPT and 3 minutes for Llama. Using a stronger computational power might remove the model size limit and allow running a local model with a quality that is similar to GPT-3.5.

\subsection{LLMs for assisted report generation}
The introduction raised the question of to what extent can (local) large language models assist forensic report writing.

Even if the use of LLMs for forensic reporting purposes brings up new opportunities, we do not believe it is yet possible to automate the generation of a complete digital forensics report, in particular the sections in which the knowledge and opinion of the investigator are predominant. Upon conducting our experiments, it is evident that there exists substantial potential to enhance the process of report writing, particularly within the introductory sections of forensic reports. Our empirical analysis underscores the efficacy of text summarization and text reduction in these segments. In such cases, the advantages outweigh any associated limitations: starting with a first draft instead of a blank page improves the report writing efficiency, one of the beneficial aspects automation can provide according to \citet{michelet2023automation}. It should be noted, however, that the text presented cannot be trusted unconditionally, and it requires careful and comprehensive examination. This caution is especially appropriate when dealing with more advanced sections in order not to fall for hallucinatory data.

Nevertheless, it is vital to recognize that we are still a considerable distance away from achieving true autonomy in report writing. The finalization of reports will invariably necessitate the discerning expertise and oversight of qualified human professionals.

\section{Limitations and Next Steps}\label{sec:limitations}

A limitation of this study arises from the use of reports prepared by students for testing and evaluation. Due to their inexperience, students likely follow the course materials (i.e., instructor suggestions), and these reports may not always accurately reflect the variety and complexity found in real-world forensic investigations. In practice, different institutions and forensic experts may use unique report formats that may not perfectly match the standard structure defined in this study. However, we assume that most institutions have a somewhat consistent structure internally in their reports.

Our experiments were conducted using general LLMs. It is important to recognize that the results obtained with these general models may show significant variation when applied to more specialized forensic scenarios, i.e., the results are likely to be significantly better on specialized models, which is planned in the future. 

Another limitation lies in the small scale of our experiments, which focused on a single case and a single example of the tested section. This limitation was necessitated by the manual nature of several steps within our experiment. This manual execution represents another potential source of human error. However, it is important to emphasize that this manual process closely mirrors real-world investigative procedures and thus simulates the authentic forensic investigation process realistically. 

\paragraph{Future work} To improve and strengthen the experiment, a wider range of models should be tested, a wider range of prompts and input data should be used, a larger number of texts should be generated, and these texts should be formally evaluated. 
One possibility would be to replace the manual querying using the web interface, by interacting with the API to automate this step.
A part of the text quality evaluation would still require manual intervention. This would provide a better overview of the differences between the models (hundreds of them can be found online, and some could even be trained or fine-tuned for forensic reporting purposes), a stronger understanding of the prompt engineering and input data formatting required to optimize the quality of texts generated, and a more precise evaluation of the generated texts. Another element that should be discussed is the standardization of the forensic report sections, which is not yet achieved. The description and discussion of these sections might change for institutes or investigators that use a different structure, and having a common report construction might help better understand each other.

\section{Conclusion}\label{sec:conclusion}
In this study, we began by exploring Large Language Models for automating forensic report generation, an important but often overlooked phase in digital forensics investigations. While automation has made advances in various forensic tasks, report generation has remained relatively untouched due to its complexity.

Our research has provided valuable insights, including an analysis of report structure, experimental results showing the feasibility of text generation, and discussions of challenges with (Local) Large Language Models. We found that certain report sections can be automated, although the quality of the generated text varies by model. However, issues such as model size, generation time, and hallucinations were identified as challenges.

This first step is encouraging, but it is only the beginning. Future research should incorporate a broader range of LLMs, diverse data sources, and formal text evaluations. In addition, standardization of forensic reporting structures across institutions is essential for consistency.

\section*{Acknowledgements}
We would like to thank Francesco Servida and Timothy Bollé who created this fictive case, along with the DFRWS EU reviewers whose comments greatly helped improve the paper.

\section*{CRediT authorship contribution statement}
\textbf{Gaëtan Michelet:} Conceptualization, Methodology, Investigation, Writing - Original Draft, Writing - Review \& Editing, Visualization.
\textbf{Frank Breitinger:} Conceptualization,  Methodology, Investigation, Writing - Review \& Editing, Supervision.

\section*{Declaration of interest}
The authors declare that they have no known competing financial interests or personal relationships that could have appeared to influence the work reported in this paper.


\bibliographystyle{model5-names.bst}



\appendix

\input{appendix}


\end{document}

%% file: appendix.tex
\section{Input data examples}\label{annex:input_data_examples}

\begin{figure*}
\scriptsize 
\begin{verbatim}
05.02.2019 12:16(UTC+0) Direction:Incoming
Wonder Woman created the group Secret
Fichier source: USERDATA (ExtX)/Root/data/org.telegram.messenger/files/cache4.db : 0x8DE5E (Table: messages,Taille : 720896 octets)

05.02.2019 12:16(UTC+0)Direction:Incoming, 695862679 (Wonder Woman)
Han, Obi Wan, This mission will be tricky and we need the best!
Fichier source: USERDATA (ExtX)/Root/data/org.telegram.messenger/files/cache4.db : 0x8DDE4 (Table: messages,Taille : 720896 octets)
USERDATA (ExtX)/Root/data/org.telegram.messenger/files/cache4.db-wal : 0xE72EF (Table: users,Taille : 4144752 octets)

05.02.2019 12:17(UTC+0)Direction:Entrant, 695862679 (Wonder Woman)
A package is waiting for you at the Krystal on E colonial Drive near Union Park.
Fichier source: USERDATA (ExtX)/Root/data/org.telegram.messenger/files/cache4.db : 0x8DD56 (Table: messages,Taille : 720896 octets)
USERDATA (ExtX)/Root/data/org.telegram.messenger/files/cache4.db-wal : 0xE72EF (Table: users,Taille : 4144752 octets)
\end{verbatim}
    \caption{Several messages extracted from the telegram chat group described in the previous Figure, from the Cellebrite report (copy of tool report).}
    \label{fig:ufed_chat_messages}
\end{figure*}

\begin{figure*}
\scriptsize 
\begin{verbatim}

From 	                     Timestamp: Time               Deleted        Instant Message	     Status
Body 
Source file information
                                         
-                          05.02.2019 07:16:04(UTC-5)    -               -                   -     
Wonder Woman created the group Secret	
USERDATA(ExtX)/Root/data/org.telegram.messenger/files/cache4.db : 0x8DE5E (Table: messages, Size: 720896 bytes)
                           
695862679 Wonder Woman	    05.02.2019 07:16:54(UTC-5)	   -               -	                   -
Han, Obi Wan, This mission will be tricky and we need the best!	
USERDATA (ExtX)/Root/data/org.telegram.messenger/files/cache4.db : 0x8DDE4 (Table: messages, Size: 720896 bytes) & 
USERDATA(ExtX)/Root/data/org.telegram.messenger/files/cache4.db-wal : 0xE72EF (Table: users, Size: 4144752 bytes)
		  
695862679 Wonder Woman     05.02.2019 07:17:49(UTC-5)	    -              -	                   -
A package is waiting for you at the Krystal on E colonial Drive near Union Park.
USERDATA (ExtX)/Root/data/org.telegram.messenger/files/cache4.db : 0x8DD56 (Table: messages, Size: 720896 bytes) & 
USERDATA(ExtX)/Root/data/org.telegram.messenger/files/cache4.db-wal : 0xE72EF (Table: users, Size: 4144752 bytes)
\end{verbatim}
    \caption{Same messages as in Fig.~\ref{fig:ufed_chat_messages} but formatted for better readability in the lab log (copy of lab log).}
    \label{fig:log_chat_messages}
\end{figure*}

\begin{figure*}
\scriptsize 
\begin{verbatim}
From                   ; Body                                                                             ; Timestamp: Time
                       ; Wonder Woman created the group Secret                                            ; 05.02.2019 07:16:04(UTC-5)
695862679 Wonder Woman ; Han, Obi Wan,                                                                    ; 05.02.2019 07:16:54(UTC-5)
                       ; This mission will be tricky and we need the best!                                ;
695862679 Wonder Woman ; A package is waiting for you at the Krystal on E colonial Drive near Union Park. ; 05.02.2019 07:17:49(UTC-5)
\end{verbatim}
    \caption{Same messages as in Fig.~\ref{fig:log_chat_messages}, filtered, and formatted in csv for better readability (CSV-formatted data).}
    \label{fig:csv_chat_messages}
\end{figure*}

\newpage

\section{The Mandate}\label{app:mandate}

\includepdf[pages=-]{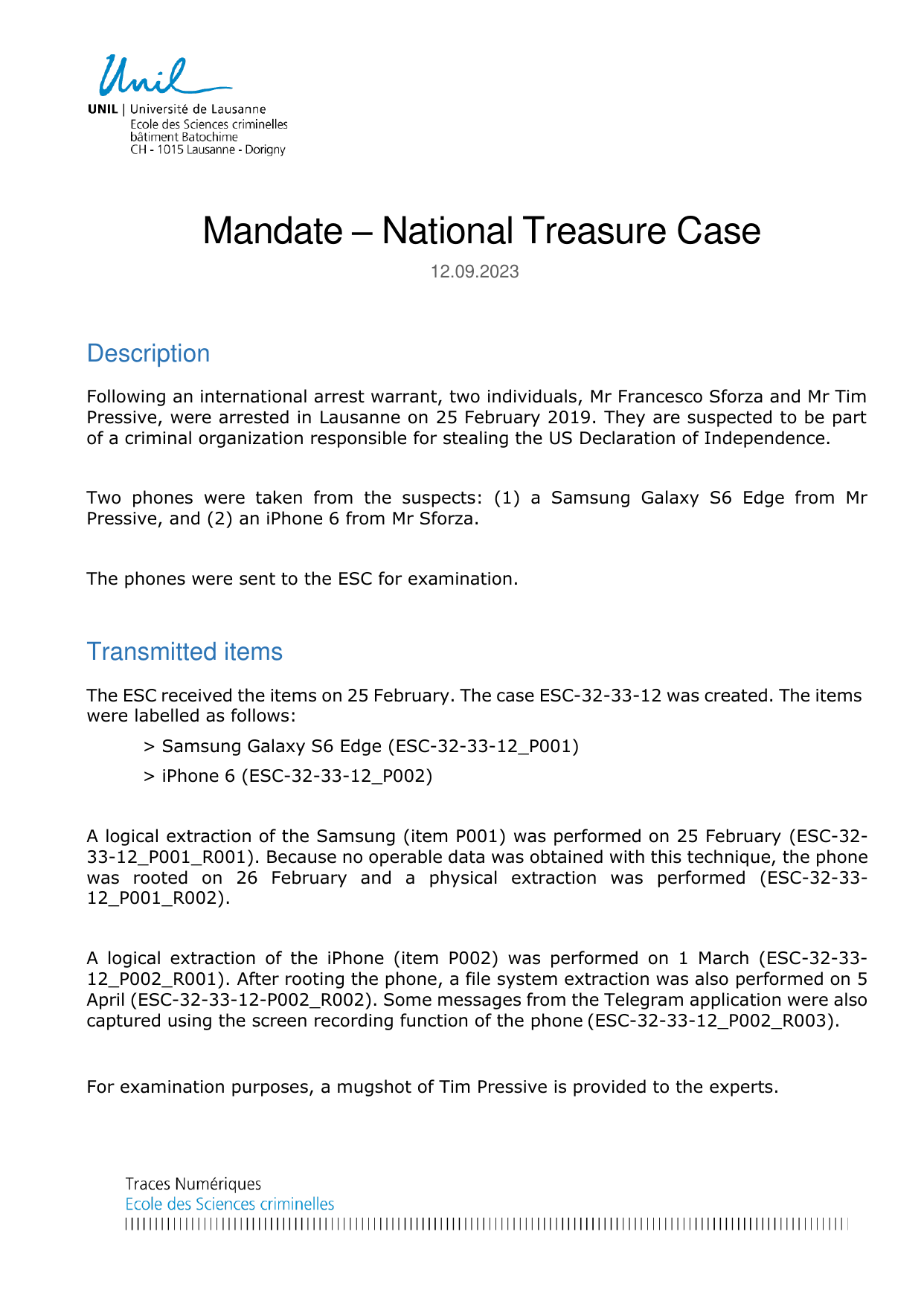}